\documentclass[12pt,preprint]{aastex}

\shorttitle{Influence of turbulence models}
\shortauthors{Jacoutot et al.}

\begin{document}

\title{Numerical simulation of excitation of solar oscillation modes \\
    for different turbulent models}
\author{L. Jacoutot
 }
\affil{Center for Turbulence Research, Stanford, CA 94305, USA}
\email{jacoutot@stanford.edu}
\author{A. G. Kosovichev
}
\affil{Hansen Experimental Physics Laboratory, Stanford, CA 94305, USA}
\and
\author{A. Wray
 and N. N. Mansour
 }
\affil{NASA Ames Research Center, CA, USA }

    
\begin{abstract}
The goal of this research is to investigate how well various turbulence models can describe physical properties of the upper convective boundary layer of the Sun. An accurate modeling of the turbulence motions is necessary for understanding the excitation mechanisms of solar oscillation modes. We have carried out realistic numerical simulations 
using several different physical Large Eddy Simulation (LES) models (Hyperviscosity approach, Smagorinsky, and dynamic models) to investigate how the differences in turbulence modeling affect the damping and excitation of the oscillations and their spectral properties and compare with observations. We have first calculated the oscillation power spectra of radial and non-radial modes supported by the computational box with the different turbulence models.  Then we have calculated the work integral input to the modes to estimate the influence of the turbulence model on the depth and strength of the oscillation sources. We have compared these results with previous studies and with the observed properties of solar oscillations. We find that the dynamic turbulence model provides the best agreement with the helioseismic observations.
\end{abstract}


\section{Introduction}

Dominant acoustic sources within the Sun are generated by strong fluctuations in the outer convective layers. Turbulent motions stochastically excite the resonant modes via Reynolds stresses and entropy fluctuations. The dominant driving comes from the interaction of the nonadiabatic, incoherent pressure fluctuations with the coherent mode displacement \citep{Nordlund2001}.  The modes excitation sources occur close to the surface, mainly in the intergranular lanes and near the boundaries of granules \citep{Stein2001}. Thus an accurate modeling of the turbulence motions is necessary to understand the excitation mechanisms of solar oscillation modes. The correct choice of turbulence model is also important in many other astrophysical simulations. 

The objective of this research is to study the influence of turbulence models on the excitation mechanisms by means of realistic numerical simulations. We have compared different physical Large Eddy Simulation (LES) models (Hyperviscosity approach, Smagorinsky, and dynamic models) to show the influence on the damping and excitation of the oscillations.  

The organization of this paper is as follows. In \S2, we describe the main lines of the code and the different turbulence models. The kinetic energy of the radial modes obtained with the different turbulence models are presented in \S 3. Then a comparison of the results obtained with the different turbulence models for the non-radial modes is given in \S 4. The work integral input to the modes is calculated in \S5 in order to estimate the influence of the turbulence models on the depth of the oscillation sources.

\section{Numerical model}
We use a 3D, compressible, non-linear radiative-hydrodynamics code developed by Dr. A. Wray for simulating the upper solar photosphere and lower atmosphere. This code takes into account several physical phenomena: compressible fluid flow in a highly stratified medium, radiative energy transfer between the fluid elements, and a real-gas equation of state. The equations we solve are the grid-cell average (henceforth called ``average'') conservation of mass, 
\begin{equation}
\frac{\partial \rho}{ \partial t}+\left(\rho u_i\right)_{,i}=0,
\end{equation}
conservation of momentum,
\begin{equation}
\frac{\partial \rho u_i}{ \partial t}+\left(\rho u_i u_j + (P_{ij}+\rho\tau_{ij})\right)_{,j}=-\rho\phi_{,i},
\end{equation}
and conservation of energy,
\begin{equation}
\frac{\partial E}{ \partial t}+\left(E u_i + (P_{ij}+\rho\tau_{ij})u_j-(\kappa+\kappa_T) T_{,i} + F_i^{rad}\right)_{,i}=0,
\end{equation}
where $\rho$ denotes the average mass density, $u_i$ the Favre-averaged velocity, and $E$ the average  total energy density $E=\frac{1}{2}\rho u_i u_i + \rho e + \rho \phi$, where $\phi$ is the gravitational potential and $e$ is the Favre-averaged internal energy density per unit mass. $F_i^{rad}$ is the radiative flux, calculated by solution of the radiative transfer equation, and $P_{ij}$ is the average stress tensor $P_{ij}=\left(p+{2}\mu u_{k,k}/{3}\right)\delta{ij}-\mu\left(u_{i,j}+u_{j,i}\right)$ with $\mu$ the viscosity. The fluid pressure $p$ is a function of $e$ and $\rho$ through a tabulated equation of state;  $\tau_{ij}$ is the Reynolds stress, $\kappa$ is the molecular thermal conductivity, and $\kappa_T$ is the turbulent thermal conductivity.

The turbulence models used are the original Smagorinsky model \citep{smago} and its dynamic formulation \citep{Germano}, herein called simply the dynamic model. The Reynolds stresses $\tau_{ij}$  are modeled as in the usual Samgorinsky formulation by writing them in terms of an eddy viscosity formed from the large-scale stress tensor $S_{ij}\equiv(u_{i,j}+u_{j,i})/2$:\begin{equation}
\tau_{ij}=-2C_S\Delta^2|S|(S_{ij}-u_{k,k}\delta_{ij}/3)+2C_C\Delta^2|S|^2\delta_{ij}/3,
\end{equation}
where $|S|\equiv \sqrt{2S_{ij}S_{ij}}$ and $\Delta \equiv (\Delta x \Delta y \Delta z)^{1/3}$, $\Delta x$, $\Delta y$, and $\Delta z$ are the grid step sizes. The two parameters, $C_S$ and $C_C$, which are the classical Smagorinsky coefficient as used in incompressible flow and a coefficient associated with the trace of the subgrid Reynolds stress (which is absent in incompressible flow), must be specified in some way.  Constant values were used in some runs and in others these parameters were determined by the dynamic method using planar averages. The turbulent Prandtl number was taken as unity to set $\kappa_T$. The molecular viscosity $\mu$ and thermal conductivity $\kappa$ were taken to be zero as their solar values are exceedingly small.

We simulate the upper layers of the convection zone using 66x66x40 grid cells. The region extends 6x6~Mm horizontally and from 2.5~Mm below the visible surface to 0.5 Mm above the surface. This computational box have been chosen to directly compare with the previous results obtained by \citet{Stein2001} who used a numerical viscosity model not related to a particular turbulence model.

\section{Kinetic energy of radial modes}
\begin{figure}
\begin{center}
\begin{tabular}{cc}
    \includegraphics[width=0.45\textwidth]{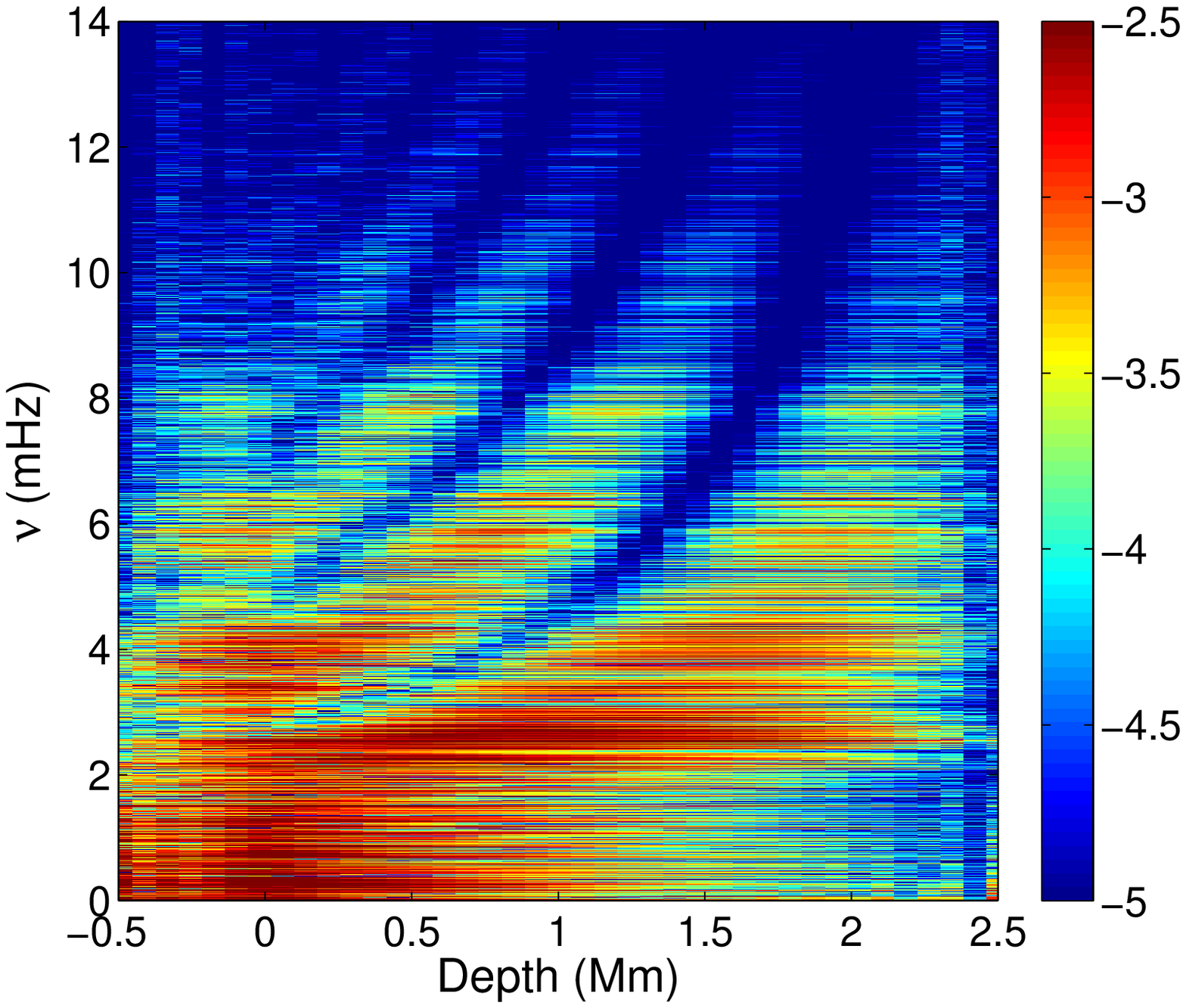}& \includegraphics[width=0.45\textwidth]{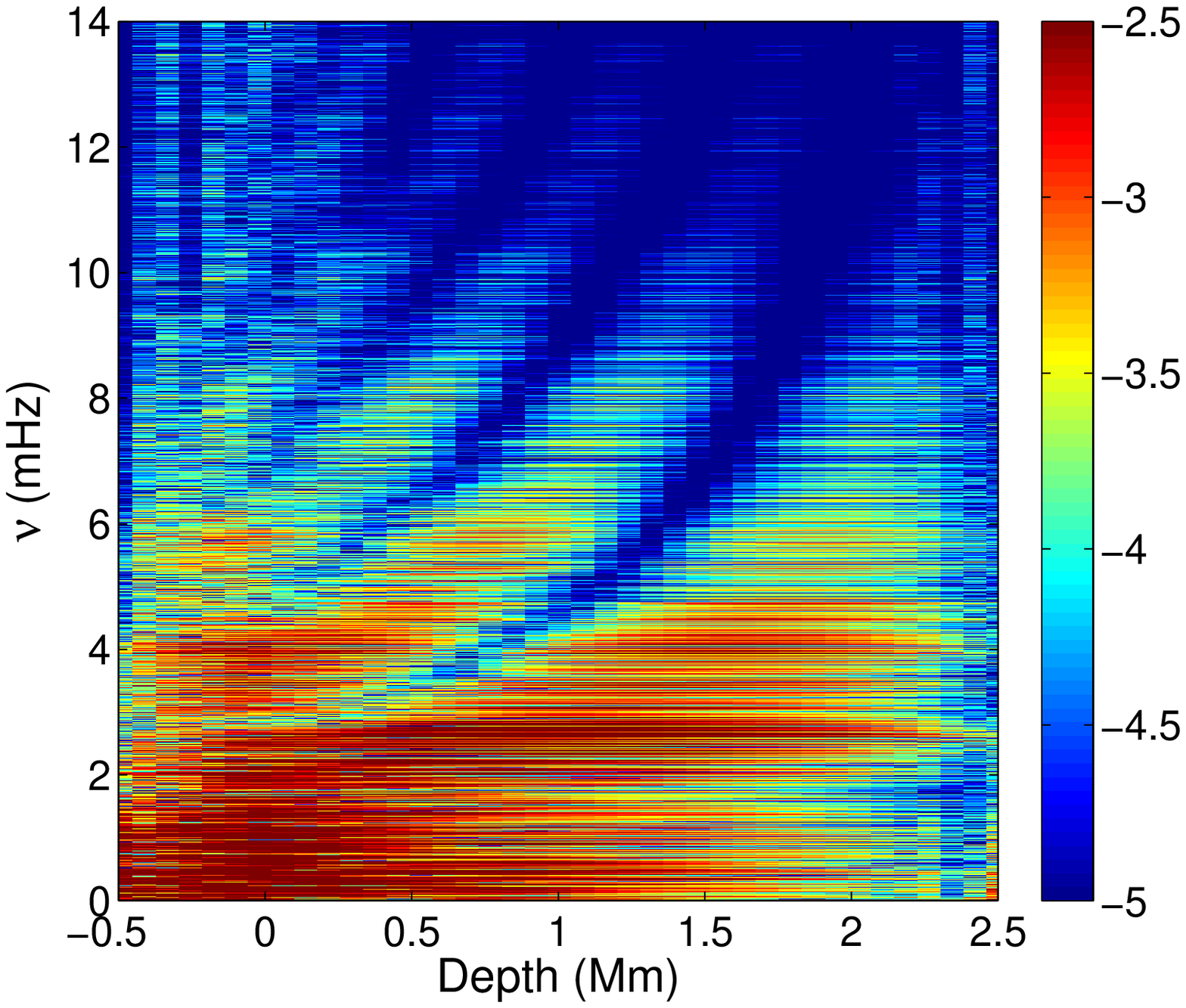} \\
  \includegraphics[width=0.45\textwidth]{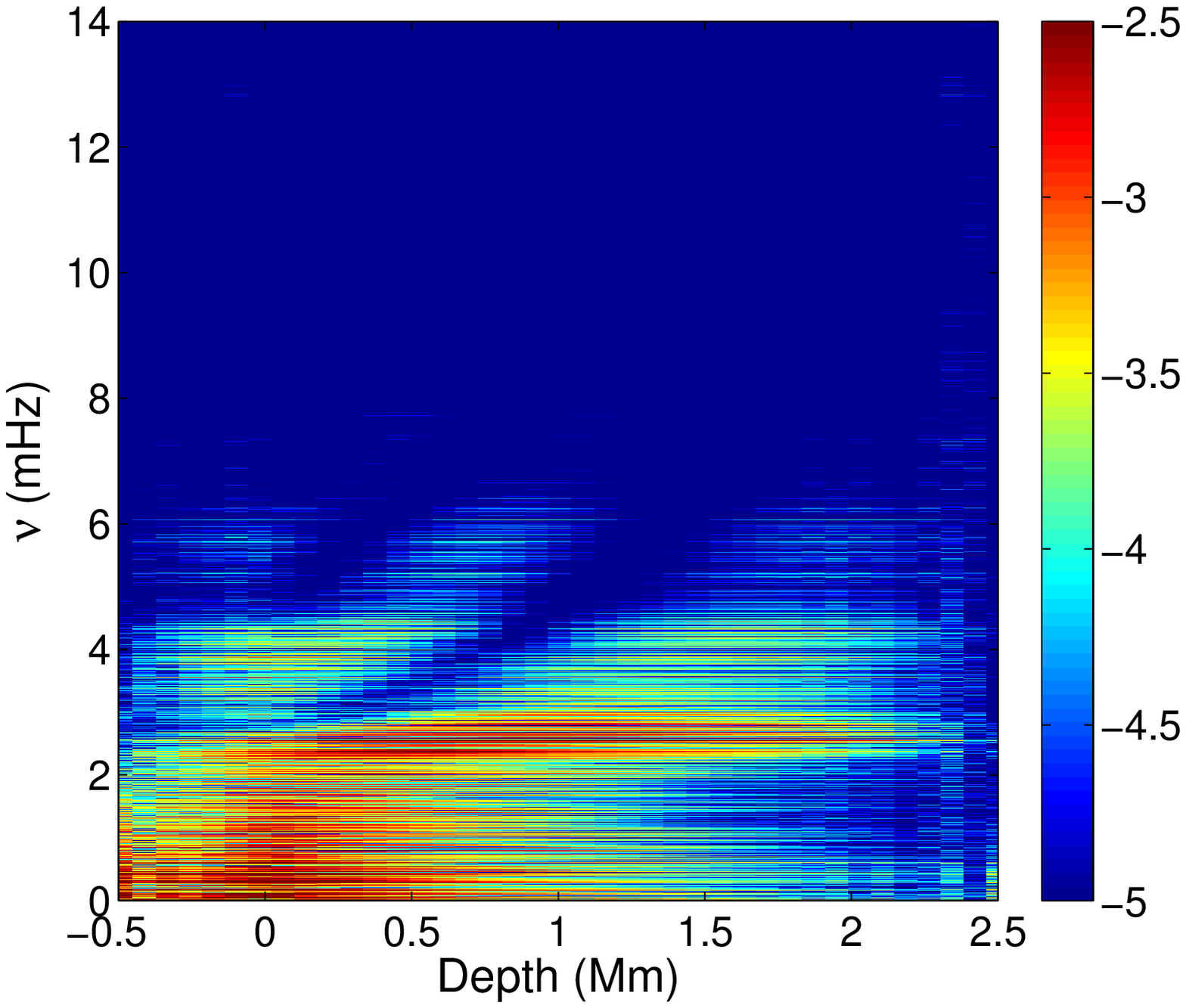}& \includegraphics[width=0.45\textwidth]{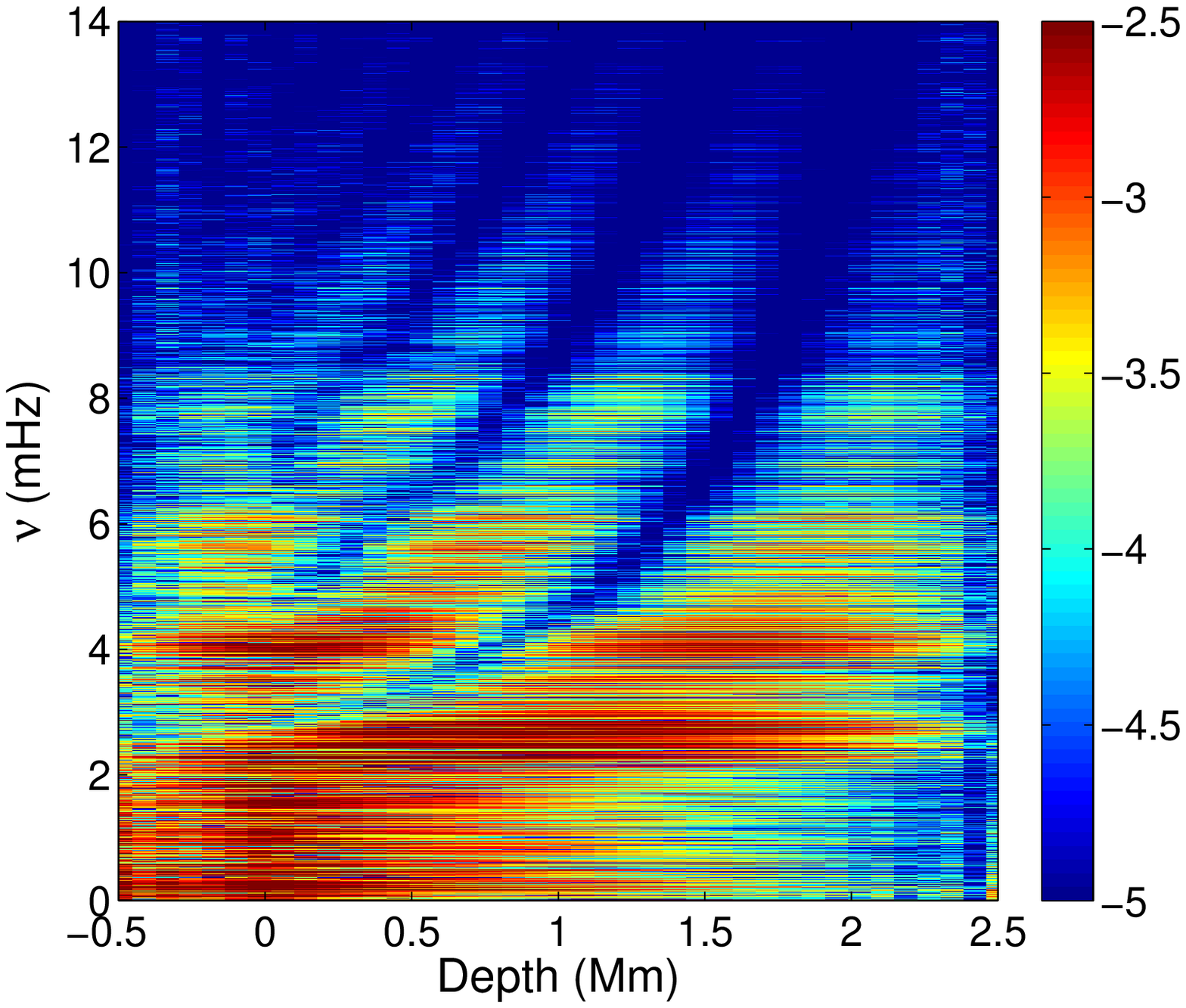} \\
\end{tabular}
\caption{Logarithm of kinetic energy as a function of depth and frequency ($erg.cm^{-3}$). \textit{Top left:} Minimal hyperviscosity approach. \textit{Top right:} Enhanced hyperviscosity approach. \textit{Bottom left:} Smagorinsky model ($C_S$=0.2). \textit{Bottom right:} Dynamic model. 
 \label{rhoE_w}}
\end{center}
\end{figure}

\begin{figure}
\begin{center}
\begin{tabular}{cc}
\includegraphics[width=0.45\textwidth]{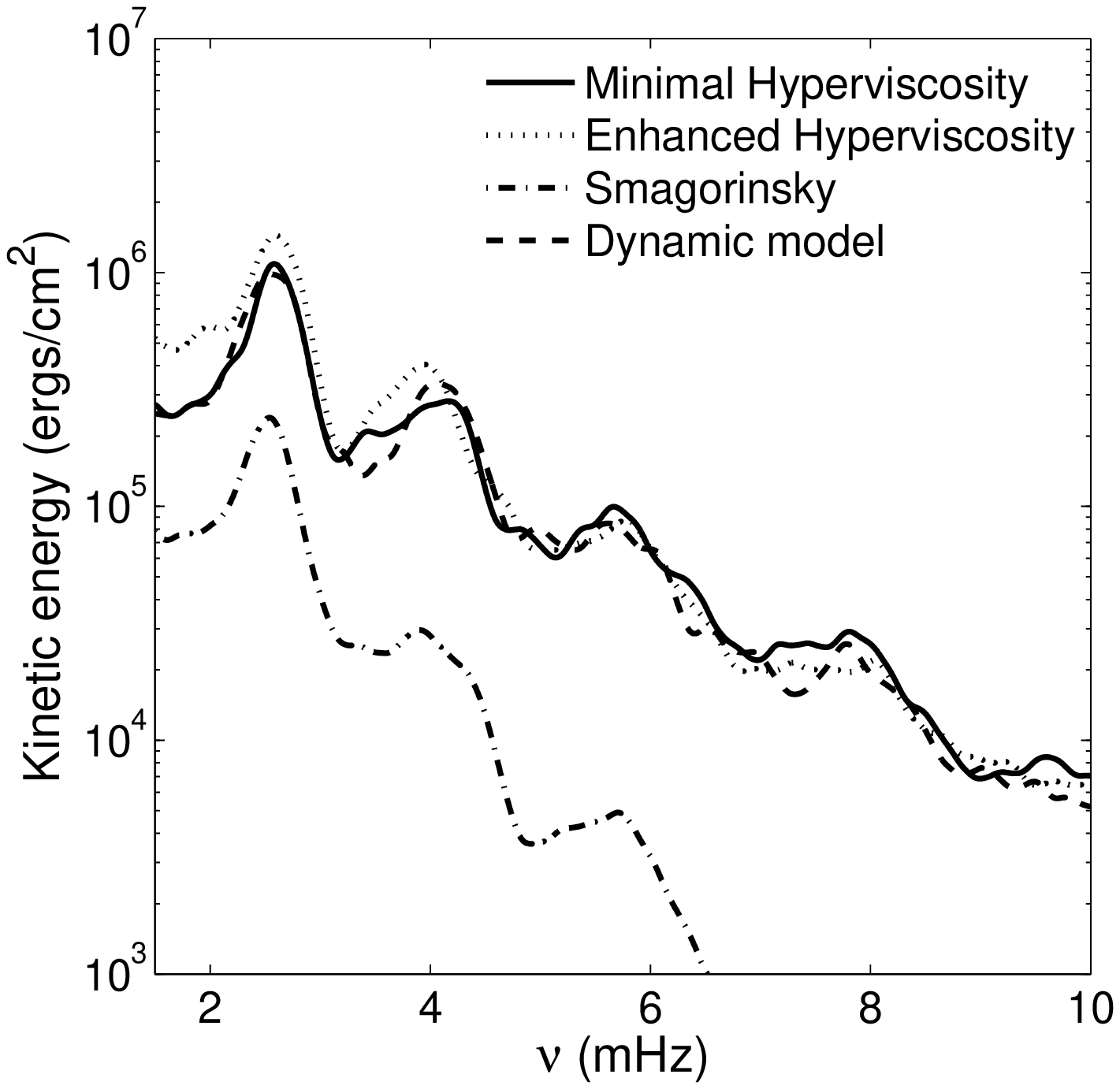} &
\includegraphics[width=0.45\textwidth]{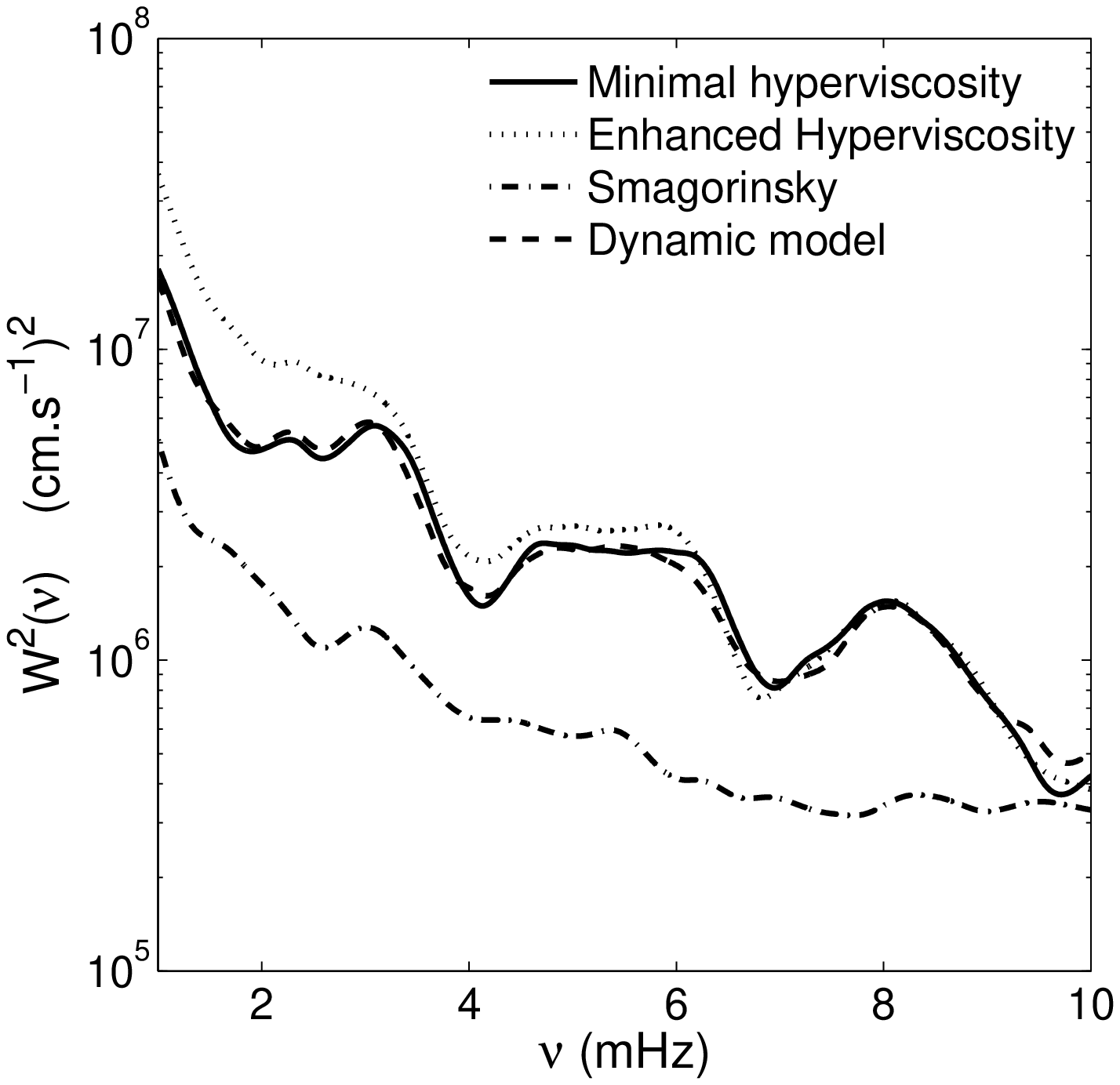}\\
 \end{tabular}
\caption{\textit{Left:} Logarithm of the kinetic energy, horizontally averaged and integrated over depth for the different turbulence models $l=0$ ($erg.cm^{-2}$). \textit{Right:} Power spectra density of vertical velocity with different turbulent modes for angular degree $l=740$ at 0.5 Mm below the surface ($(cm.s^{-1})^2$).
 \label{int_rhoE_w}}
 \end{center}
\end{figure}

First we have studied how the kinetic energy is dissipated for the different turbulence models. Specifically we have calculated the oscillation power spectra of radial modes. Those modes are extracted by horizontal averaging of the vertical velocity and Fourier transform in time. The results presented here (Figure \ref{rhoE_w} and Figure \ref{int_rhoE_w}) have been obtained with simulations of 60 hours of solar time, and the snapshots were saved every 30 seconds. 

Three modes can be clearly seen in the spectra of the horizontally averaged, depth-integrated kinetic energy obtained with all the three turbulence models (Figure \ref{int_rhoE_w} (left panel)), they correspond to the maxima in the kinetic energy. The resonant frequencies supported by the computational box are 2.6, 4.0, 5.6 mHz. These frequencies are very close to the values obtained by \citet{Stein2001}. We see that for the hyperviscosity approach and the dynamic model the modes within the computation box are excited with the same magnitude whereas the Smagorinsky model yields a lower magnitude.

The kinetic energy spectra as a function of frequency and depth (Figure \ref{rhoE_w}) confirm that the dissipation is weaker with the minimal hyperviscosity approach. In this case, almost nearly the numerical dissipation plays a role. Thus the kinetic energy is higher for high frequencies in comparison with the results obtained with the other turbulence models. The spectrum obtained with the enhanced hyperviscosity approach and the dynamic model show a weakly higher dissipation compared to the calculation with the minimal hyperviscosity approach. Moreover the dissipation applied by the hyperviscosity approach and the dynamic model does not interact with the three oscillations modes. In fact, the dissipation scale is smaller than the scale of the acoustic modes. With the Smagorinsky model, the excitation of the modes is weaker and the scale of the dissipation is close to the scale of the third mode. The patterns do not extend above 6 mHz (against 12 mHz with the hyperviscosity approach and 10 mHz with the dynamic model). The cut-off frequency obtained with the Smagorinsky model is very similar of that obtained by \citet{Stein2001}: the kinetic energy is very low above 6 mHz. Their calculations have been performed with artificial numerical viscosity.

These results show that the spectral kinetic energy of the radial modes is rather similar for the hyperviscosity and dynamic model, but the Smagorinsky model is too much dissipative.

\section{Kinetic energy of non-radial modes}
The non-radial modes have been extracted by performing 2d spatial Fourier transform on the surfaces of vertical velocity at each time step. Thus we obtain a power spectrum for a particular horizontal wave number $k_h^2=k_x^2+k_y^2$. Then we take Fourier transform in time. 
We have especially considered modes with horizontal wavelength corresponding to the box size L~=~6~Mm.
 That corresponds to angular degree of $l\simeq740$ ($k_h=1~Mm^{-1}$). Figure \ref{int_rhoE_w} (right panel) shows the power spectra for the different turbulent models. The leftmost peak corresponds to the surface gravity (f) mode, while the others correspond to acoustic (p) modes. The f-mode peak has the same frequency independent of the depth of the computational box (2.8 mHz). One can see that the Smagorinsky model gives lowest power, and the modes are less excited. The hyperviscosity approach and the dynamic model give very close results. The modes have approximately the same magnitude. The influence of the turbulence models is similar for the radial and non-radial modes. 

\section{Calculation of the p-mode excitation rates}
The mode excitation rate is calculated using the same method presented by \citet{Stein2001}. The rate of energy input to the modes per unit surface area ($erg.cm^{-2}.s^{-1}$) is
\begin{equation} 
\label{eqn:rate}
\frac{\Delta <E_{\omega}>}{\Delta t}=\frac{\omega^2 |\int_r dr \delta P_{\omega}^*  \left( \partial \xi_{\omega} / \partial r \right)|^2}{8\Delta \nu E_{\omega}},
\end{equation}
where $\delta P_{\omega}^*$ is the Fourier transform of the nonadiabatic total pressure. $\Delta \nu$ the frequency interval for the Fourier transform. $\xi_{\omega}$ is the mode displacement for the radial mode of angular frequency $\omega$. It is obtained from the eigenmode calculations of \citet{Chrsit1996}. His spherically symmetric model S gives 35 radial modes what provide much better frequency coverage in comparison with the three resonant modes obtained within the simulation box. $E_{\omega}$ is the mode energy per unit surface area ($erg.cm^{-2}$) defined as: 
\begin{equation} 
\label{eqn:Ew}
E_{\omega}=\frac{1}{2} \omega^2 \int_r dr \rho {\xi_{\omega}}^2 {\left(\frac{r}{R} \right)}^2.
\end{equation}

\begin{figure}
\begin{center}
 \includegraphics[width=0.5\textwidth]{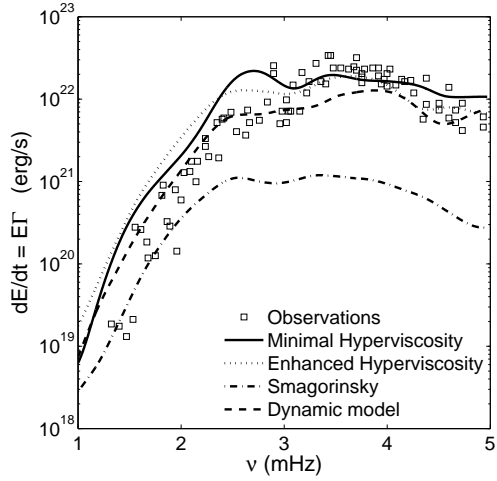}
\caption{Comparison of observed and calculated rate of stochastic energy input to modes for the entire solar surface ($erg.s^{-1}$). Observed distributions ($squares$) come from SOHO GOLF for $l=0-3$ \citep{Cortes1999} \label{rate}}.
\end{center}
\end{figure}

We have calculated the rate of stochastic energy input to modes for the entire solar surface in order to compare with the observed results (Figure \ref{rate}). The rate of energy input to the solar modes is obtained by multiplying the rate of the simulation modes by the ratio of the mode mass of the solar modes over the mode mass in the computational domain. The comparison between observed and calculated rate shows that the Smagorinsky model gives too low values (more than one decade below the observations). When we consider the minimal hyperviscosity approach the main discrepancy is located between 2 and 3 mHz. We can see a peak in this range, which is much less pronounced in the observed results. This peak is present also with the enhanced hyperviscosity approach but it is lower. It is no longer present with the dynamic model. The best agreement is obtained below 3 mHz using the dynamic model. Above 3 mHz, both the enhanced hyperviscosity approach and dynamic model provide good results. The origin of the peak or plateau in the solar power spectrum in terms of the mixing length theory was discussed by \citet{Gough} and \citet{Balmforth}.

\begin{figure}
\begin{center}
\begin{tabular}{cc}
    \includegraphics[width=0.45\textwidth]{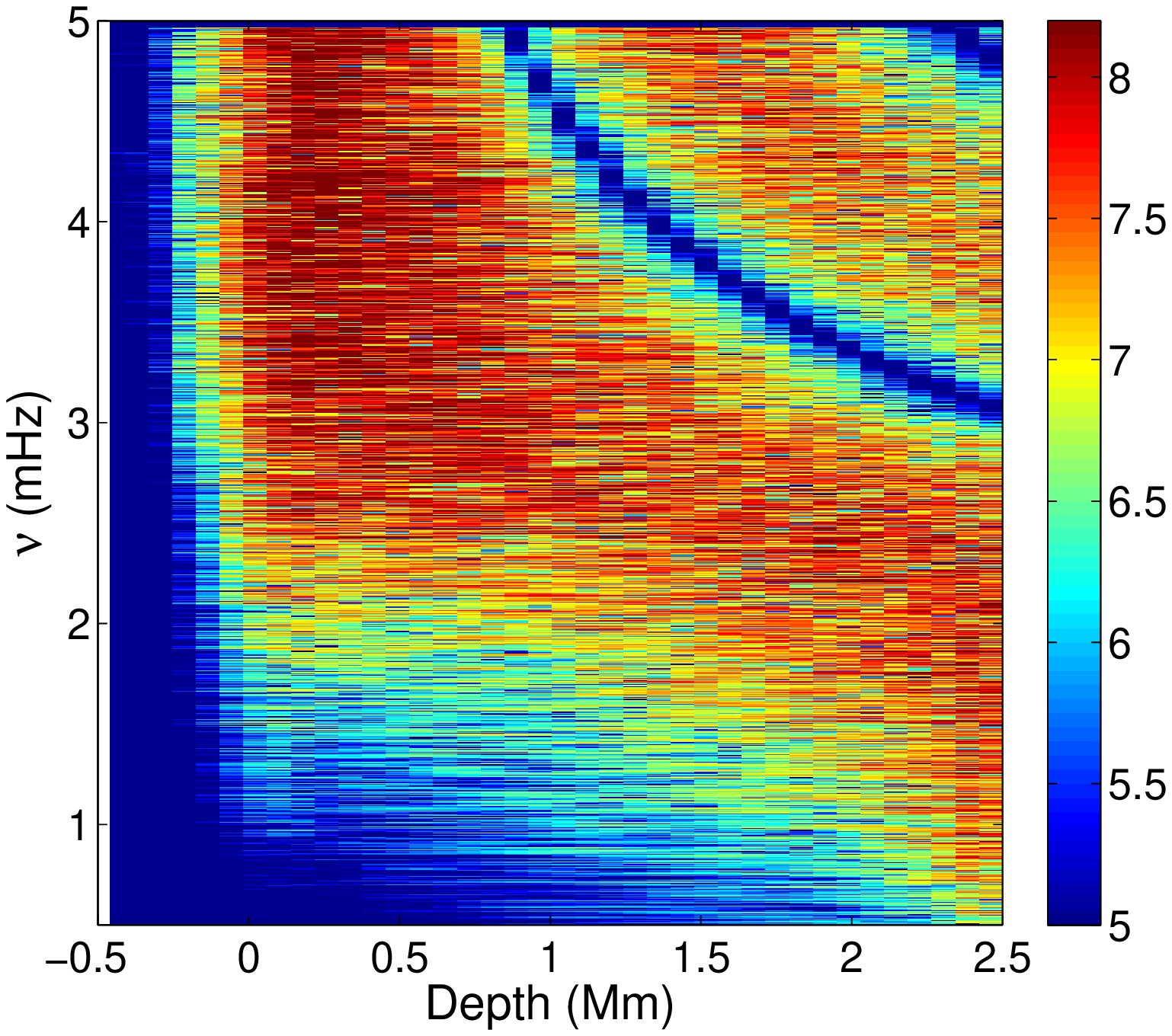}& \includegraphics[width=0.45\textwidth]{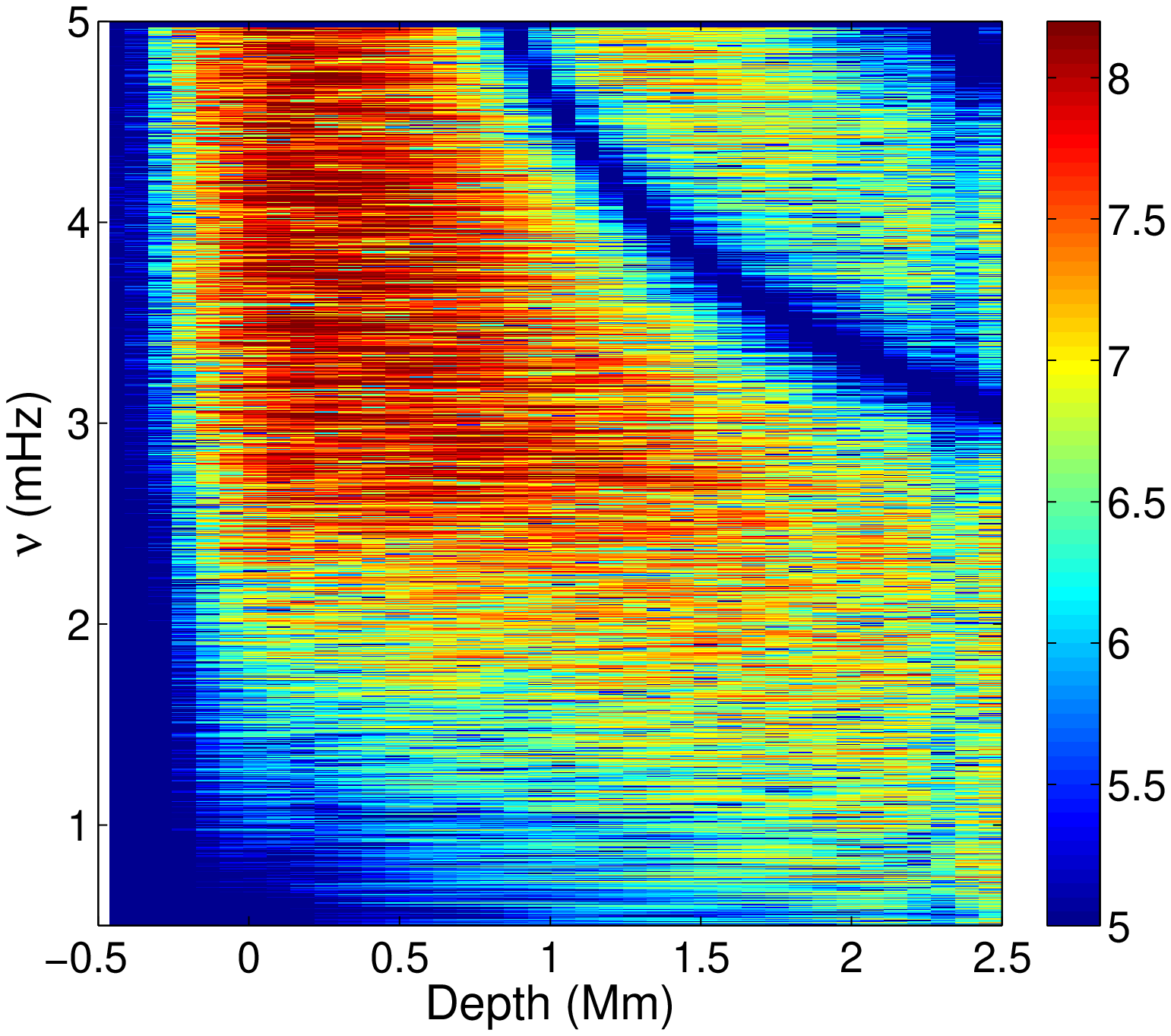} \\
  \includegraphics[width=0.45\textwidth]{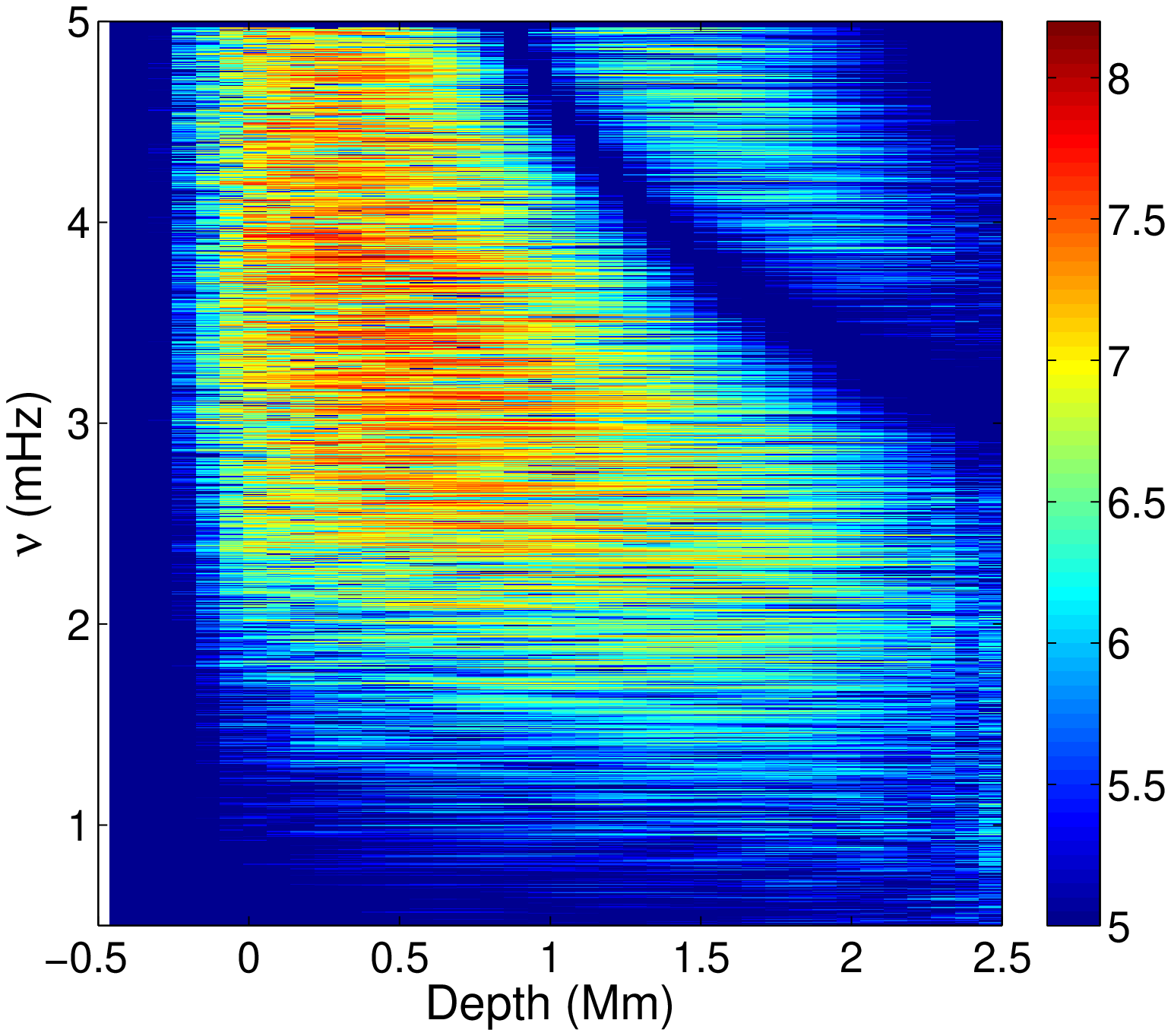}& \includegraphics[width=0.45\textwidth]{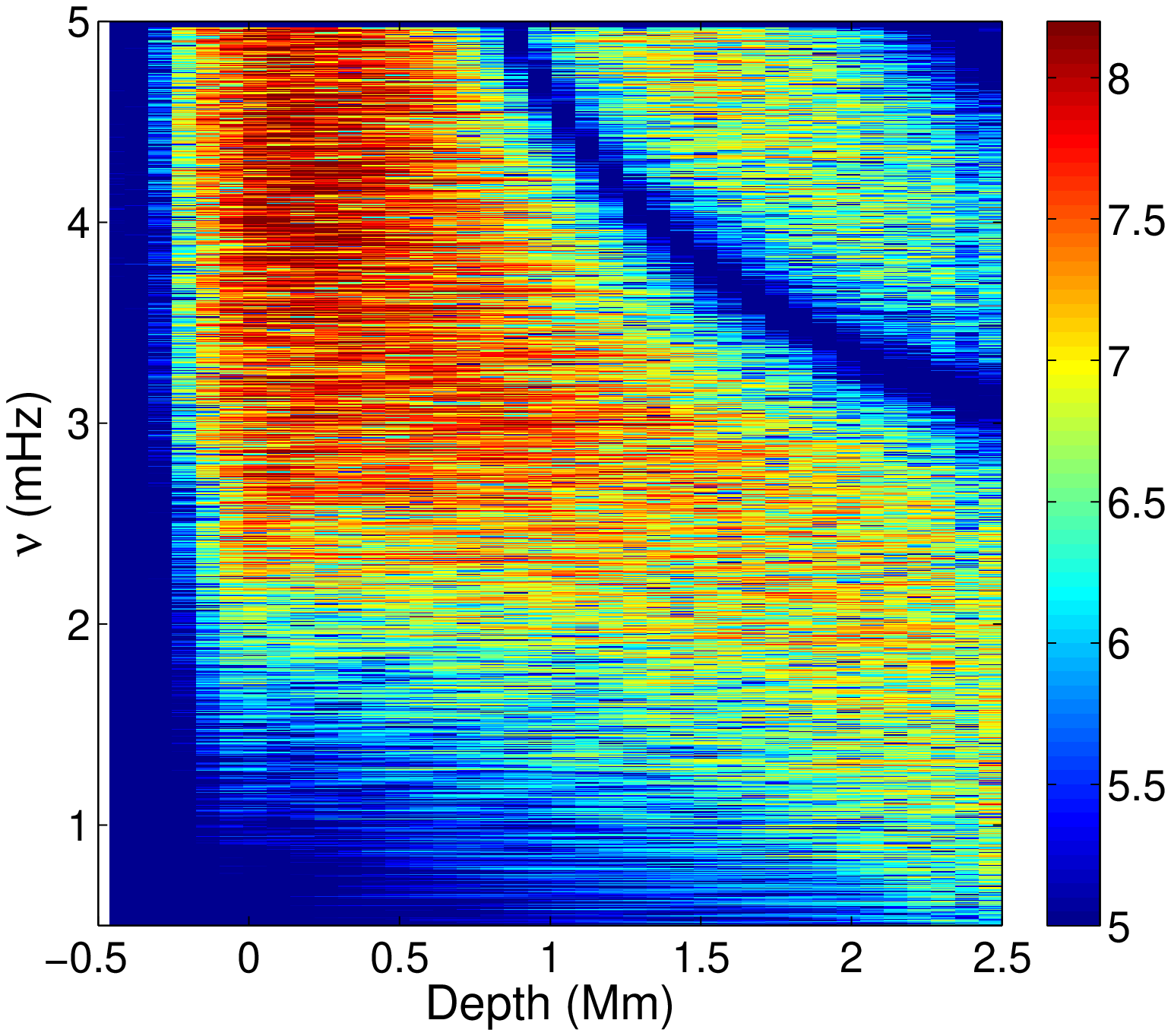} \\
\end{tabular}
\caption{Logarithm of the work integrand (eq. (\ref{eqn:rate}) in units of $erg.cm^{-2}.s^{-1}$), as a function of depth and frequency. \textit{Top left:} Minimal hyperviscosity approach. \textit{Top right:} Enhanced hyperviscosity approach (A=0.4). \textit{Bottom left:} Smagorinsky model ($C_S$=0.2). \textit{Bottom right:} Dynamic model. 
 \label{integrand}}
\end{center}
\end{figure}

The distributions of the integrand of the work integral as a function of depth and frequency (Figure \ref{integrand}) can explain the presence of this peak. The distributions are similar to the results obtained by \citet{Stein2001}. The most driving is concentrated between the surface and 500 km depth around 3-4 mHz. We can see that the main differences between the magnitude obtained with the different turbulence models are located in the region around 2 Mm. We can observe that the excitation obtained with the minimal hyperviscosity (Figure \ref{integrand} (top left panel)) is high between 2 and 3 mHz at each vertical position. Conversely the excitation decreases as the depth increases with the enhanced hyperviscosity approach and the dynamic model. The excitation magnitude becomes low when the depth is greater than 2 Mm with the enhanced hyperviscosity approach (1.5 Mm with the dynamic model). These differences explain the difference in magnitude of the peak between 2 and 3 mHz for the rate of stochastic energy. The presence of the peak in the observations underlines that the excitation is only located close to the visible surface of the Sun. 

\section{Conclusion}
The goals of this research was to investigate how well various turbulence models can describe the convective properties of the upper boundary layer of the Sun and to study the excitation and damping of acoustic oscillations. Results obtained with the hyperviscosity approach have been compared with those obtained with the Smagorinsky and dynamic turbulence models. We have seen that the dissipation is very high with the Smagorinsky model while the hyperviscosity approach and dynamic modes give similar results. Besides we find that the dynamic turbulence model provides the best agreement with observations.

\newpage

\clearpage

\end{document}